\documentclass[preprint,superscriptaddress]{revtex4-1}

\usepackage{array}
\usepackage{subcaption}
\usepackage{hyperref}
\usepackage{threeparttable}
\usepackage{rotating}

\usepackage{filecontents}
\newcolumntype{P}[1]{>{\centering\arraybackslash}p{#1}}
\newcolumntype{M}[1]{>{\centering\arraybackslash}m{#1}}

\makeatletter
\newcommand{\mypm}{\mathbin{\mathpalette\@mypm\relax}}
\newcommand{\@mypm}[2]{\ooalign{%
  \raisebox{.1\height}{$#1+$}\cr
  \smash{\raisebox{-.6\height}{$#1-$}}\cr}}
\makeatother

\usepackage[version=3]{mhchem}
\usepackage[T1]{fontenc}
\usepackage{graphicx}
\usepackage{natbib}
\usepackage{url}
\usepackage{amsmath}
\usepackage{amssymb}
\usepackage{mathrsfs}
\usepackage{color}
\usepackage{soul,xcolor}
\usepackage{comment}
\usepackage[english]{babel}
\emergencystretch=\maxdimen
\begin{document}

\title{Liquid-liquid displacement in slippery liquid-infused membranes (SLIMs)}

\author{Hanieh Bazyar}
\affiliation{Department of science and technology, Soft matter, Fluidics and Interfaces (SFI), University of Twente, Enschede, The Netherlands}
\affiliation{Wetsus, European centre of excellence for sustainable water technology, Leeuwarden, The Netherlands}

\author{Pengyu Lv}
\affiliation{Department of Science and Technology, Physics of Fluids (POF), Max Planck - University of Twente Center for Complex Fluid Dynamics, University of Twente, Enschede, The Netherlands}

\author{Jeffery A. Wood}
\affiliation{Department of science and technology, Soft matter, Fluidics and Interfaces (SFI), University of Twente, Enschede, The Netherlands}

\author{Slawomir Porada}
\affiliation{Department of science and technology, Soft matter, Fluidics and Interfaces (SFI), University of Twente, Enschede, The Netherlands}
\affiliation{Wetsus, European centre of excellence for sustainable water technology, Leeuwarden, The Netherlands}

\author{Detlef Lohse}
\affiliation{Department of Science and Technology, Physics of Fluids (POF), Max Planck - University of Twente Center for Complex Fluid Dynamics, University of Twente, Enschede, The Netherlands}

\author{Rob G. H. Lammertink}
\email{r.g.h.lammertink@utwente.nl}
\affiliation{Department of science and technology, Soft matter, Fluidics and Interfaces (SFI), University of Twente, Enschede, The Netherlands}

\begin{abstract}
Liquid-infused membranes inspired by slippery liquid-infused porous surfaces (SLIPS) have been recently introduced to membrane technology. The gating mechanism of these membranes is expected to give rise to anti-fouling properties and multi-phase transport capabilities. However, the long-term retention of the infusion liquid has not yet been explored. To address this issue, we investigate the retention of the infusion liquid in slippery liquid-infused membranes (SLIMs) via liquid-liquid displacement porometry (LLDP) experiments combined with microscopic observations of the displacement mechanism. Our results reveal that pores will be opened corresponding to the capillary pressure, leading to preferential flow pathways for water transport. The LLDP results further suggest the presence of liquid-lined pores in SLIM. This hypothesis is analyzed theoretically using an interfacial pore flow model. We find that the displacement patterns correspond to capillary fingering in immiscible displacement in porous media. The related physics regarding two-phase flow in porous media is used to confirm the permeation mechanism appearing in SLIMs. In order to experimentally observe liquid-liquid displacement, a microfluidic chip mimicking a porous medium is designed and a highly ramified structure with trapped infusion liquid is observed. The remaining infusion liquid is retained as pools, bridges and thin films around pillar structures in the chip, which further confirms liquid-lining. Fractal dimension analysis, along with evaluation of the fluid (non-wetting phase) saturation, further confirms that the fractal patterns correspond to capillary fingering, which is consistent with an invasion percolation with trapping (IPT) model.
\end{abstract}

\maketitle

Bio-inspired interfacial materials with non-wetting properties have broad technological implications for areas ranging from biomedical devices and fuel transport to architecture \cite{quere2008wetting}. Lotus-leaf inspired superhydrophobic surfaces are well-known, owing to properties such as drag reduction, anti-icing, anti-frosting and self-cleaning \cite{Darmanin2014,JianyongLV2014,doi:10.1021/acs.langmuir.5b04754,doi:10.1021/acs.langmuir.6b00064,doi:10.1021/acsami.5b07881,doi:10.1021/la5021143,doi:10.1021/acsami.6b01133}. However, superhydrophobic surfaces are prone to failure due to elevated pressures and temperatures and by dissolution of the trapped air into the surrounding fluid \cite{doi:10.1021/acs.langmuir.5b04754,doi:10.1021/am404077h}. This is particularly significant for low surface tension liquids \cite{doi:10.1021/la103097y}. Recently, a novel class of functional surfaces known as slippery liquid-infused porous surfaces (SLIPS) has been introduced \cite{Wong2011Bioinspired}. Inspired by Nepenthes pitcher plants \cite{Bohn28092004}, the nano/microstructured substrate is used to lock-in an intermediary liquid. This liquid, stabilized by capillary forces, forms a smooth, low hysteresis lubrication layer which is responsible for the non-wetting properties \cite{doi:10.1021/nn302310q,837566,Epstein14082012}. In comparison with superhydrophobic surfaces, SLIPS can potentially improve anti-icing or suppress frost accretion \cite{doi:10.1021/nn302310q,doi:10.1021/acs.langmuir.6b00064,doi:10.1021/nn303372q}, operate at high pressures and temperatures \cite{Wong2011Bioinspired,doi:10.1021/am404077h}, and possibly reduce drag \cite{doi:10.1021/la5021143,fu_arenas_leonardi_hultmark_2017}. Self-healing by capillary wicking, repelling a variety of liquids, and anti-biofouling are other potentially advantageous properties of SLIPS \cite{Epstein14082012,Wong2011Bioinspired}. Recently, it has been shown that the capillary-stabilized liquid in a membrane pore can form a reconfigurable gate which can selectively let fluids pass through. This so-called gating mechanism gives SLIPS the capability of multiphase transport without clogging \cite{837566}. The liquid-lined gating mechanism was previously explored via gas-liquid porometry where the remaining liquid film thickness on the pore wall was estimated experimentally and theoretically \cite{ADMI:ADMI201600025}.

In previous studies, the fabrication methods of liquid infused surfaces were complex, time-consuming and the substrate selectivity to match solid and liquid chemistry was limited. Some methods required multi-step processing, high temperatures and drying \cite{doi:10.1021/nn303372q,doi:10.1021/nn303867y}. Using well-matched membrane-liquid combinations with low surface energies for water, e.g. membranes prepared from fluorinated polymers, avoids the extra step of solid surface energy reduction. Examples of these membranes are polyvinylidene fluoride (PVDF) and polytetrafluoroethylene (PTFE). \cite{doi:10.1021/am404077h,837566,ADMI:ADMI201600025}.

Introduction of SLIPS to membranes provides potentially anti-fouling properties, as well as pressure responsive pores for selective fluid transport \cite{837566,ADMA:ADMA201600797}. These properties are attractive for separation applications. The capability of these membranes in efficient gas-liquid sorting for three phase air-water-oil mixtures \cite{837566}, for example, is relevant for oily waste water treatment. Before the potential of slippery liquid infused membranes (SLIM) can be realized, a thorough understanding of the retention of the infusion liquid under dynamic conditions of transport of the immiscible fluid is required.

Two-phase flow or immiscible displacement in any porous media, such as porous membranes, is typically governed by viscous and capillary forces, characterized by two dimensionless numbers, the viscosity ratio

\begin{equation}
M = \frac{\mu_i}{\mu_d},\ \label{eq5}
\end{equation}

\noindent and the capillary number

\begin{equation}
Ca = \frac{\mu_i v}{\gamma}.\ \label{eq6}
\end{equation}

\noindent Here $\mu_i$ and $\mu_d$ are the viscosities of invading (water) and defending (oil) fluids respectively, $v$ is the superficial velocity and $\gamma$ is the interfacial tension between the two immiscible fluids.
The domain of validity of different basic mechanisms, i.e. capillary fingering, viscous fingering, or stable displacement, can be mapped on to a phase diagram \cite{lenormand_touboul_zarcone_1988}. The displacement can also be classified according to drainage or imbibition, where the defending (displaced) or invading (displacing) fluid preferentially wets the solid surface, respectively \cite{PhysRevLett.115.164501,lenormand_touboul_zarcone_1988}. Microfluidic techniques have been used to investigate the displacement mechanisms. These techniques offer the opportunity to fabricate micromodels resembling a porous medium with regular as well as irregular pore shapes, anisotropy and pore sizes. Pore-scale multi-phase displacement phenomena have been experimentally observed in pore network patterns fabricated in materials such as silicon \cite{WRCR:WRCR9508,doi:10.1021/ef101732k}, glass \cite{PhysRevE.70.016303}, PDMS \cite{arXiv:0909.0758,PhysRevE.84.026311,B802373P,PhysRevE.82.046315}, polyester and thiolene-based resin \cite{0953-8984-2-S-008,lenormand_touboul_zarcone_1988,PhysRevLett.119.208005}.

Liquid-gated membranes are expected to possess anti-fouling characteristics and multi-phase transport capabilities \cite{837566}. However, knowledge regarding the retention of the infusion liquid within these membranes is still limited. To address this issue, we report on retention of the infusion liquid in SLIM and microscopic observation of the displacement mechanism. Liquid-liquid displacement porometry (LLDP) is done in a flux-controlled mode by pushing pure water through PVDF membranes infused with perfluoropolyether oil (Krytox 101). The results are further analyzed theoretically using an interfacial pore flow model. Finally, a microfluidic chip resembling the porous medium has been used to microscopically investigate the displacement mechanism under identical capillary number and viscosity ratio. The observed flow pattern is characterized using fractal dimension analysis and fluid (non-wetting phase) saturation, methods known from porous media analysis.
Here, the permeation through SLIMs is related to two-phase flow in porous media to confirm the observed displacement mechanisms that are crucial for their promising membrane applications.

\section{Results and discussion} \label{s:Resdiscussion}
\subsection{Membrane experimental results} \label{subs:MembraneExpRes}
Liquid-infused membranes are made by infusing different types of liquids (fluorinated or hydrocarbon) into dry PVDF membranes (see section \hyperref[s:matmethods]{Materials and Methods}). The physical properties of the used liquids are shown in Table \ref{tab:1} (see SI for more information on the measurement methods).

\begin{sidewaystable}[!p]
\small
\centering
\captionsetup{justification=centering}
\caption{Physical properties of different infusion liquids} \label{tab:1}
\setlength{\tabcolsep}{1mm}{
\begin{tabular}{l l M{3 cm} M{4 cm} M{4 cm} M{3 cm}}
\hline
Infusion liquids & Chemical structure & Surface tension [$mN/m$] & Absolute viscosity at 20$^{\circ}$C [$mPa$ $s$] & Kinematic viscosity at 20$^{\circ}$C [$mm^2/s$] & Density at 24$^{\circ}$C [$g/cm^{-3}$]\\
\hline
Fluorinert FC-43 & \text{Perfluorocarbon (PFC)} & 16.3$\mypm$0.05 & 5.6$\mypm$0.08 & 3.0 & 1.88$\mypm$0.01\\
Galpore & \text{Perfluoropolyether (PFPE)} & 15.5$\mypm$0.05 & 8.6$\mypm$0.08 & 4.8 & 1.83$\mypm$0.01\\
Krytox GPL oil 101 & \text{Perfluoropolyether (PFPE)} & 16.3$\mypm$0.13 & 25$\mypm$0.1 & 13.5 & 1.85$\mypm$0.01\\
Silicone oil AR20 & \text{Polyphenylmethylsiloxane} & 21.8$\mypm$0.03 & 19.6$\mypm$0.1 & 19.4 & 1.00$\mypm$0.002\\
\hline
\end{tabular}}
\end{sidewaystable}

\begin{figure}
\centering
\includegraphics[width=0.8\linewidth]{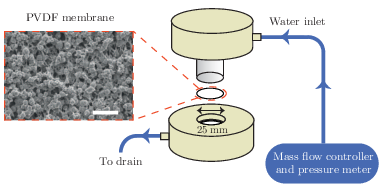}
\caption{Schematic illustration of the membrane experiment and SEM image of the PVDF membrane (Scale bar is 10 $\mu$m).}
\label{fig:Membrane_set_up_3}
\end{figure}

All liquid-infused membranes display hydrophobic behavior with a static contact angle of about 120$^{\circ}$ and contact angle hysteresis $<5^{\circ}$ using standard contact angle Goniometer (the water contact angle values on liquid infused PVDF membranes are shown in Table S1). These are the key design parameters, since a high contact angle and a low contact angle hysteresis are desirable for high droplet mobility \cite{C6SM00920D}. Liquid-liquid displacement experiments are done successively in a flux controlled mode by pushing pure water (displacing fluid) through the liquid infused membrane. The schematic illustration of the set-up is shown in Figure \ref{fig:Membrane_set_up_3}.

\begin{figure}
\centering
\includegraphics[width=0.7\linewidth]{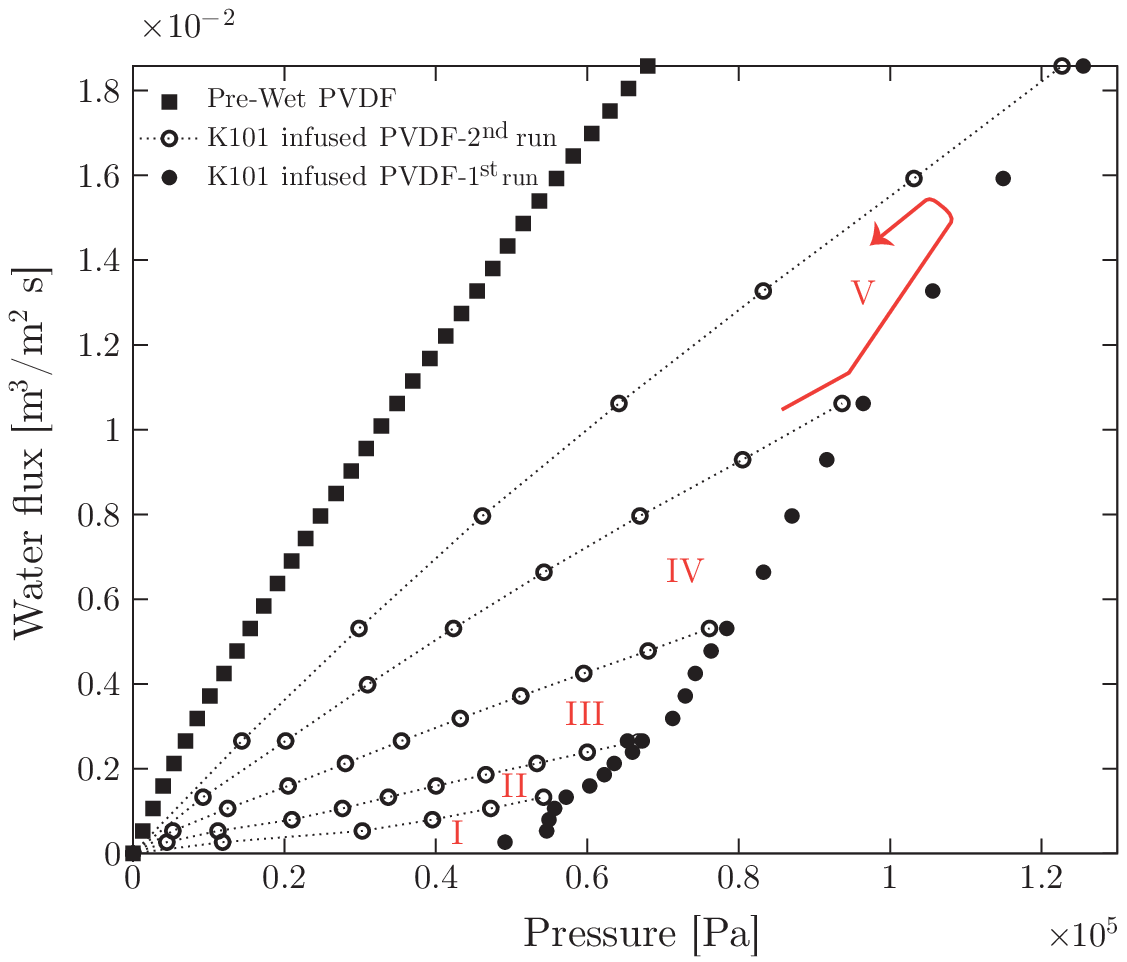}
\caption{Successive liquid-liquid displacement porometry (LLDP) experiment in a flux controlled mode on Krytox 101 infused PVDF membrane. LLDP is done in five different cycles (I-V) and each cycle consists of a 1$^\text{st}$ (filled symbols) and a 2$^\text{nd}$ run (open symbols). In each run flux is increased step-wise from zero up to a certain maximum value and pressure is measured correspondingly. A typical sequence of the measurement is shown for cycle V.}
\label{fig:Successive_flux_pressure}
\end{figure}

The result of LLDP on infused PVDF membrane with perfluoropolyether oil (Krytox 101) is shown in Figure \ref{fig:Successive_flux_pressure}. The results of other liquid infused membranes are shown in Figure S4. The LLDP experiment is performed in five different cycles. Each cycle is carried out in a flux-increasing mode, which is done twice starting from zero up to a certain flux value, i.e. the cycle consists of two runs (1$^\text{st}$ run and 2$^\text{nd}$ run). In each run, the flux is increased step-wise and pressure is measured simultaneously. At each step, flux is kept constant for 100 s and the pressure is reported for the last 40 s of each step.

In each cycle, a certain number of pores have been opened in the 1$^\text{st}$ run corresponding to the Laplace pressure

\begin{equation}
\Delta P = \frac{2\gamma \lvert \cos \theta_{E} \rvert}{r}, \label{eq1}
\end{equation}

\noindent leading to preferential flow pathways.

Here $\Delta P$ is the transmembrane pressure, $r$ is the pore radius, $\gamma$ is the interfacial tension between displacing and displaced fluid, and $\theta_E$ is the advancing contact angle of the displacing fluid on a surface of smooth and dense PVDF (total non-wetting ($\theta_E$=180$^{\circ}$) with displacing fluid is considered for liquid-infused membranes).

The opened pores remain open during the 2$^\text{nd}$ run, confirmed by the linear relation between flux and pressure \cite{MIETTONPEUCHOT199773}. An initial critical pressure of approximately $0.57\times 10^5$ Pa is required to open the biggest pore sizes. Based on the pore size distribution of the membrane obtained from capillary flow porometer (see Figures S2), the largest pore radius is 1.77$\mypm$0.11 $\mu$m. According to Young-Laplace equation (equation \ref{eq1}) and by considering total non-wetting, the corresponding pressure to open the largest pores is $0.61\times 10^5$ Pa (see Table S3 for surface and interfacial tension values), which is in good agreement with the LLDP experiments.

In order to check for the presence of the infusion liquid in the membrane after the experiment, the obtained results are compared with the results of the pre-wetted membrane (see Figure \ref{fig:Successive_flux_pressure}). The permeability of pre-wetted as well as liquid-infused membranes is calculated using Darcy's law \cite{Darcy1856Lesfontains}

\begin{equation}
Q=\frac{\kappa A}{\mu}\ \frac{dp}{dx}.\ \label{eq2}
\end{equation}

\noindent Here and in the following, $Q$ is the volumetric flow of the permeating fluid (in this case water), $\kappa$ is permeability (L$^{2}$), $A$ is the total area of the membrane, $\mu$ is the viscosity of the displacing fluid, $\frac{dp}{dx}$ is the pressure gradient across the membrane thickness, $r$ is the mean pore radius, and $\phi$ is porosity.

A simple model is then used to relate permeability to porosity $\phi$ (see SI for derivation), namely

\begin{equation}
\kappa=\frac{\phi r^{2}}{24}.\ \label{eq3}
\end{equation}

\noindent The calculated porosity is a measure of opened and active pores for water transport in the membrane. The results of the calculated permeability and the estimated fraction of the active pores are shown in Table \ref{tab:2}.

\begin{table}
\begin{threeparttable}
\small
\centering
\caption{Comparison of permeability and fraction of active pores for five different cycles of liquid infused membrane and for the pre-wetted case.} \label{tab:2}
\begin{tabular}{M{4.0cm} M{6.0cm} M{6.0cm}}
\hline
Membrane & Permeability ($\kappa$) (Darcy$^{*}$) & Estimated fraction of active pores \\
\hline
SLIM-cycle 1 & 3.12$\times$10$^{-3}$ & 0.09\\
SLIM-cycle 2 & 5.23$\times$10$^{-3}$& 0.15\\
SLIM-cycle 3 & 9.19$\times$10$^{-3}$ & 0.26\\
SLIM-cycle 4 & 1.51$\times$10$^{-2}$ & 0.43\\
SLIM-cycle 5 & 2.01$\times$10$^{-2}$ & 0.57\\
Pre-wet & 3.54$\times$10$^{-2}$ & 1\\
\hline
\end{tabular}
\begin{tablenotes}\footnotesize
\item[*] $1$ $Darcy =0.987\times10^{-12}$ $m^{2}$
\end{tablenotes}
\end{threeparttable}
\end{table}

This fraction is the ratio between the calculated porosity for the liquid infused membrane in each cycle and that of the pre-wetted membrane. The permeability of the liquid-infused membrane increases in each cycle, revealing that new pores are opened for water transport. The lower permeability values of SLIM in comparison with pre-wetted membranes suggests incomplete removal of the infusion liquid with around 43\% of retained infusion liquid in the membrane. Direct observations using the microfluidic chip shows that 27\% of the infused liquid remained in the chip at the corresponding flux value ($1.8\times10^{-2}$ m$^3$/m$^2$ s) (see Table \ref{tab:3}).

The average pore radius of the membrane based on its pore size distribution is 1.65$\mypm$0.11 $\mu$m (Figure S3). Based on the Young-Laplace equation (equation \ref{eq1}) and by assuming total non-wetting, all the pores should be opened at pressures beyond $0.61\times 10^5$ Pa. As the used membranes are tortuous porous media with interconnected pores and areas, infused liquid can be trapped. The trapped liquid will not be removed by further flux or pressure increase and thus leads to a lower permeability compared to the pre-wet one. In order to observe the re-infusion of the opened pores, long-term experiments are performed. In these experiments the pause time between each cycle was set to 12 hr. However, re-infusion is not observed at the high flux values in which the experiments have been already conducted. This is evident from the linear flux-pressure relation, that would be different when re-infused.

The experimental results are analyzed theoretically using an interfacial pore flow model to relate the flux ($J$) to pressure ($\Delta P$), namely \cite{ANTON2014219}

\begin{equation}
J(\Delta P) = \frac{N \pi \Delta P}{8\sqrt{2\pi}\ln(S)\mu l} \int_{r_{\text{min}}}^{r_{\text{max}}} r^3 \exp\left(-\frac{1}{2}\left(\frac{\ln(r/R)}{\ln(S)}\right)^2\right)\;\mathrm{d}r.\ \label{eq4}
\end{equation}

\begin{figure}
\centering
\includegraphics[width=0.7\linewidth]{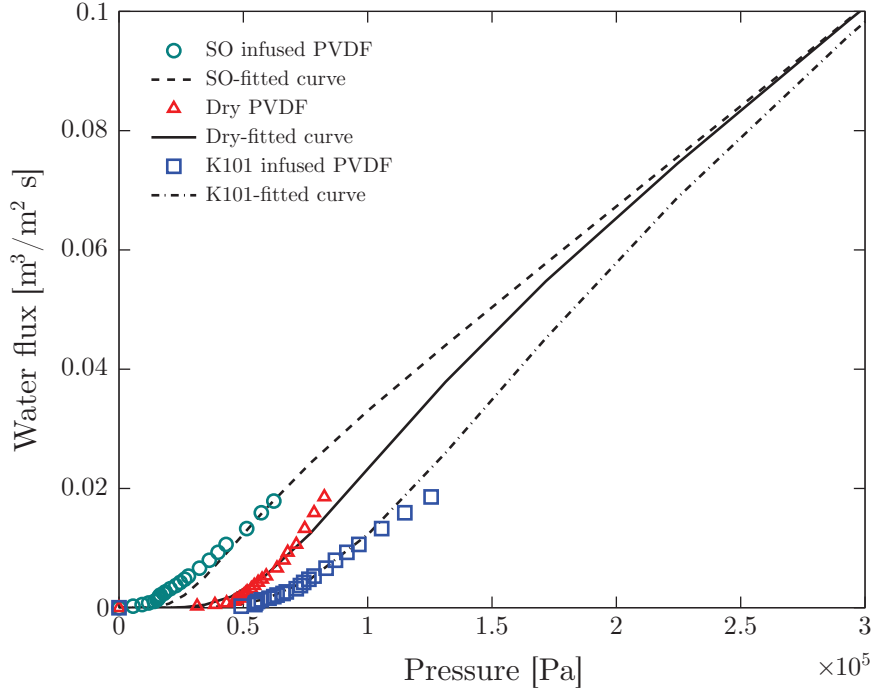}
\caption{Fitting of experimental results of LLDP to the interfacial pore flow model (equation \ref{eq4} of reference \cite{ANTON2014219}). Symbols and lines show the experimental and fitting results respectively. Fitting parameter are $N=10^7$ pore/m$^2$ (total number of pores per unit area), $R=1.65$ $\mu$m (geometric mean radius) and $S=1.2$ (geometric standard deviation)).}
\label{fig:FittingResults_thicker_new_3bar}
\end{figure}

\noindent Here $N$ is the total number of the pores per unit area, $R$ is the geometric mean radius, and $S$ is the geometric standard deviation. $J$ is the volumetric flux of displacing fluid per membrane area through the membrane, $l$ is the membrane thickness. $r_\text{{min}}$ is the equilibrium radius of curvature, i.e. Kelvin radius, which can be calculated using Young-Laplace equation (equation \ref{eq1}). $r_\text{{max}}$ is taken at least one order of magnitude larger than the largest pore size corresponding to the lowest measurable pressure by the set-up. A log-normal pore size distribution is considered for this model \cite{ANTON2014219,MORISON2008301} and was found to describe the measurements well (see SI for more details on fitting). The pore size distribution of the PVDF membrane is shown in Figure S3.

\begin{figure}
\captionsetup[subfigure]{justification=centering}
 \centering
  \begin{subfigure}[b]{0.33\textwidth}
    \centering
    \includegraphics[width=\textwidth]{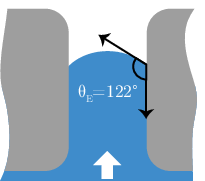}
    \caption{}
    \label{fig:Dry_pore_white}
  \end{subfigure}
  \hskip 10ex
  \begin{subfigure}[b]{0.33\textwidth}
    \centering
    \includegraphics[width=\textwidth]{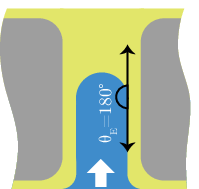}
    \caption{}
    \label{fig:SLIM_pore_white_new}
  \end{subfigure}
  \caption{Schematic illustration of water pushing through (a) dry membrane pore and (b) liquid-infused pore.}
  \label{fig:Fig4}
\end{figure}

Figure \ref{fig:FittingResults_thicker_new_3bar} shows good agreement between experimental values and fitted curves using the interfacial pore flow model (equation \ref{eq4}). It accurately describes the flux behavior in both liquid-infused and dry membranes. The result for a silicone oil (SO AR20)-infused membrane is also shown. The difference in pressure at a given flux arises purely from different Laplace pressures due to the corresponding interfacial tension values (Table S2). Since the liquid-infused membrane is preferentially wetted by the infusion liquid (this is one of the criteria for fabrication of SLIM \cite{Wong2011Bioinspired}), total non-wetting can be assumed for the invading liquid ($\theta_\text{E}$=180$^{\circ}$) (see Figure \ref{fig:Fig4}). For validation, the advancing contact angle of water on a dense membrane which was immersed in the oil was measured.  $\theta_\text{{Adv}}$=175$^{\circ}$ and $\theta_\text{{Adv}}$=160$^{\circ}$ were obtained respectively for K101 and SO AR20 (Table S1). In a dry membrane due to the absence of the wetting layer and hydrophobic character of the pore wall, partial wetting occurs. In this case, the advancing contact angle of water on smooth and dense membrane should be considered ($\theta_\text{{Adv}}$=122$^{\circ}$) (Table S1). Thus, the numerator of equation \ref{eq1} for a dry pore ($\gamma \lvert cos \theta_\text{{E}} \rvert$=38 mN/m) is between that of K101 and that of SO AR20 infused pores ($\gamma_\text{{K101-water}}=54$ mN/m and $\gamma_\text{{SO-water}}=18$ mN/m \cite{doi:10.1021/la301033h}). As it is schematically shown in Figure \ref{fig:SLIM_pore_white_new}, this further suggests the presence of remaining liquid film on the pore wall (liquid-lined pores) in SLIM.

\subsection{Microfluidic experimental results} \label{subs:MicrofluidicExpRes}
\begin{figure}
\captionsetup[subfigure]{justification=centering}
\centering
  \begin{subfigure}[b]{0.38\textwidth}
    \centering
    \includegraphics[width=\textwidth]{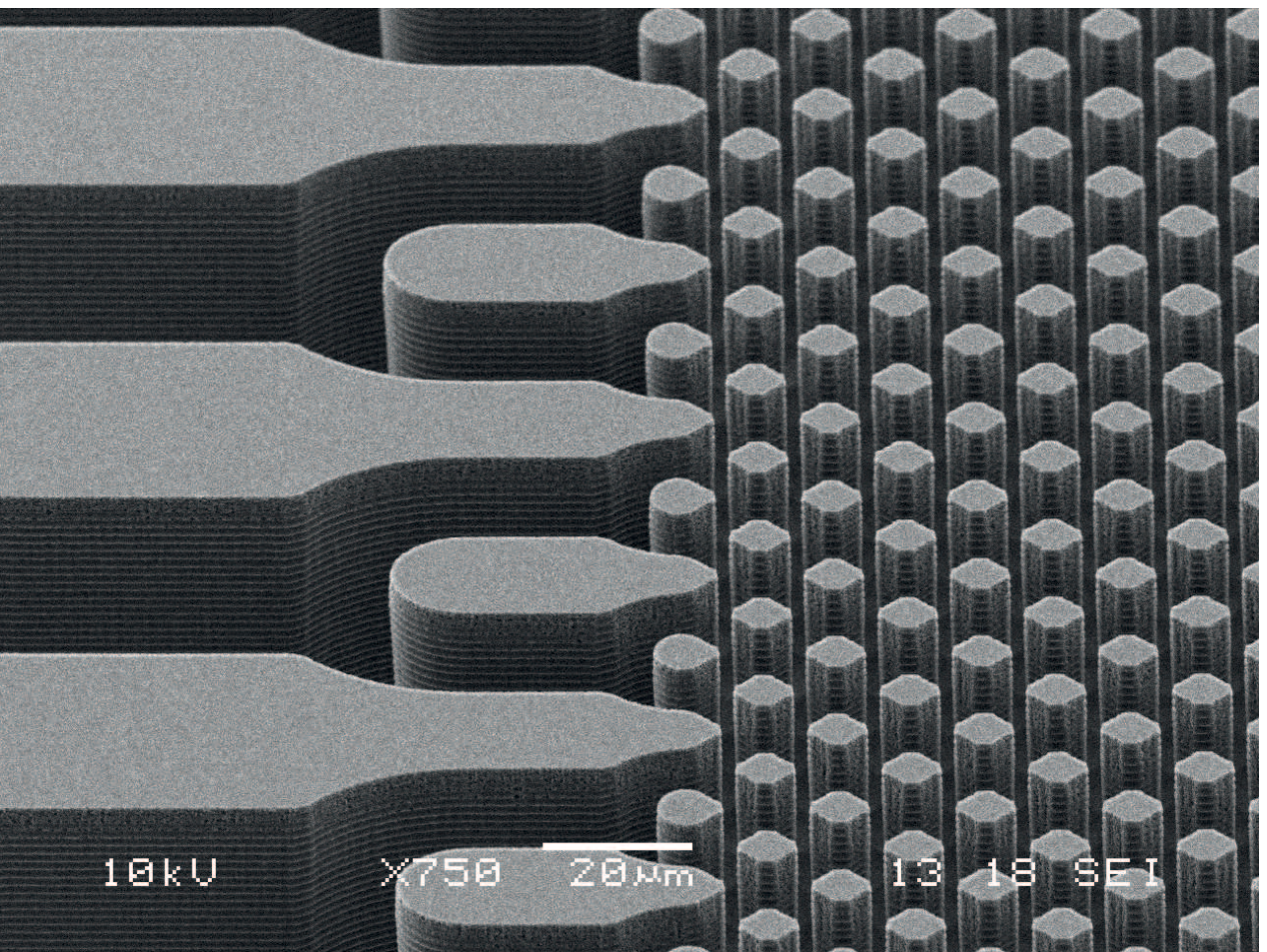}
    \caption{}
    \label{fig:Fig5_1}
  \end{subfigure}
  \hskip 2ex
  \begin{subfigure}[b]{0.38\textwidth}
    \centering
    \includegraphics[width=\textwidth]{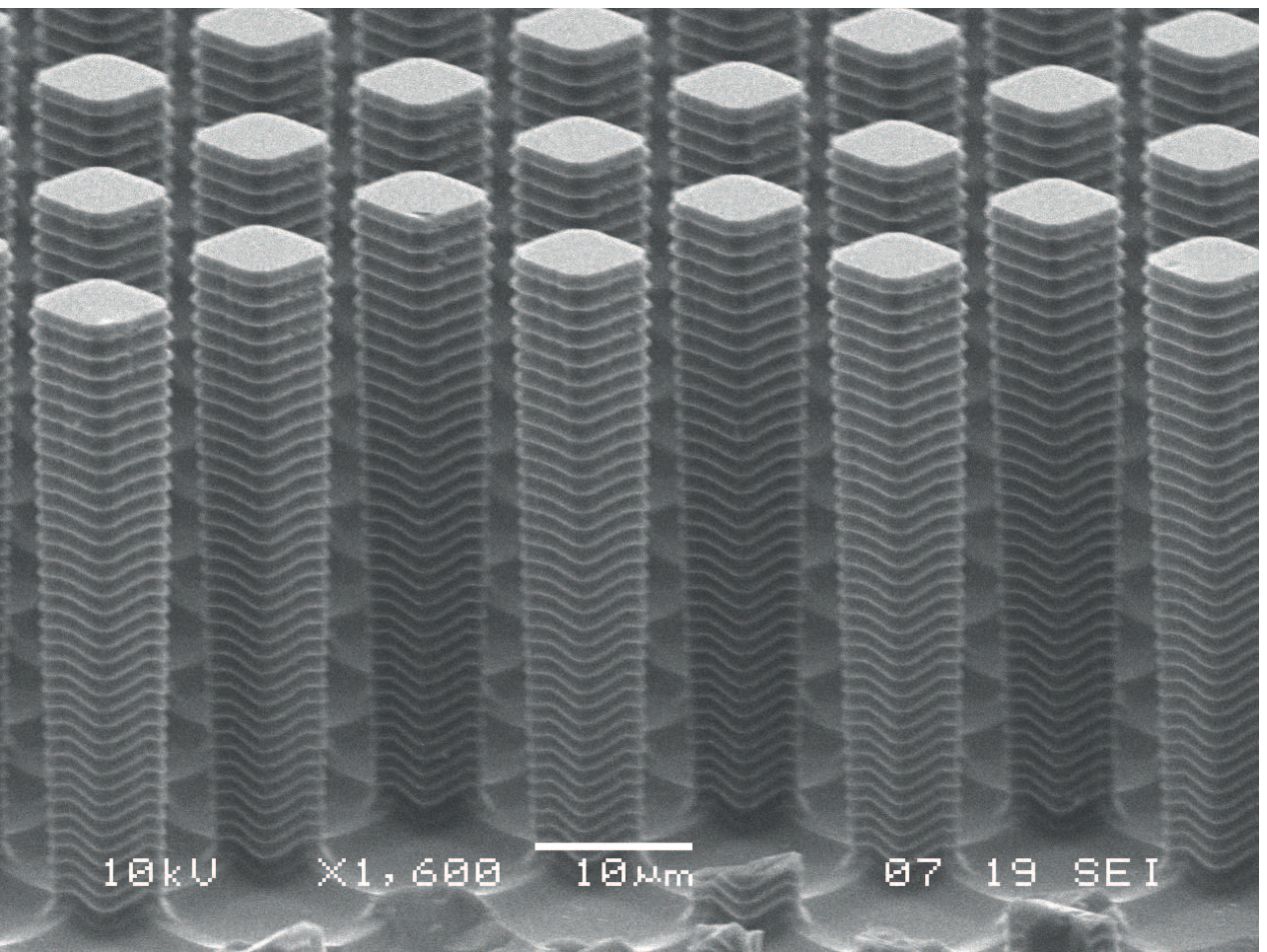}
    \caption{}
    \label{fig:Fig5_3}
  \end{subfigure}
  \caption{Scanning electron microscopy (SEM) images of (a) the inlet of the microfluidic chip, (b) pillars.}
  \label{fig:Fig5}
\end{figure}

In order to observe the displacement mechanism and the resulting flow paths, a microfluidic chip has been used as a mimic of the porous medium (Figure \ref{fig:Fig5}). The chip is fabricated in silicon using standard photolithography and reactive ion etching (see SI for detailed fabrication procedure). The pore network contains a uniform distribution of square pillars with 20 $\mu$m center to center spacing, 8 $\mu$m diameter and 50 $\mu$m height giving a porosity of 0.84.

\begin{figure}
\centering
\includegraphics[width=0.8\linewidth]{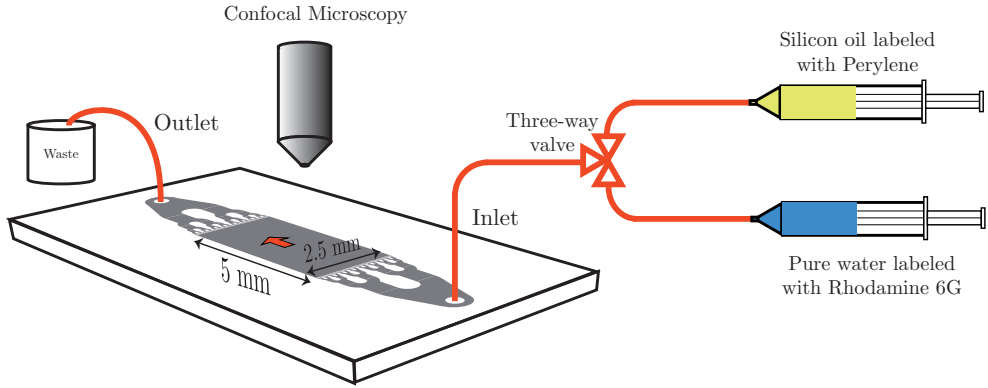}
\caption{Schematic illustration of the microfluidic experiment.}
\label{fig:Microfluidic_set_up_perspective_WithoutSEM_New}
\end{figure}

To mimic the experimental conditions of the membrane experiments, the chip is further hydrophobized with perfluorinated silane and filled with SO AR20 labeled with 8.7 mM fluorescent dye (Perylene). Water (labeled with 17.7 $\mu$M fluorescent dye (Rhodamine 6G)) is pushed through the chip from the right side using a syringe pump (Harvard apparatus PHD 2000 infuse/withdraw). The addition of the aforementioned dyes neither changed the surface tension of water nor SO AR20 considerably (see Table S3). Laser scanning confocal microscope (LSCM) is used for observation of liquid-liquid displacement. The schematic of the experimental set-up is shown in Figure \ref{fig:Microfluidic_set_up_perspective_WithoutSEM_New}.

The viscosity ratio (equation \ref{eq5}) is kept the same as in membrane experiments by using similar fluid pairs. The experiment is done at different flow rates, corresponding to the same capillary number (equation \ref{eq6}) as the membrane experiments (Table \ref{tab:3}).

\begin{figure}
\captionsetup[subfigure]{justification=centering}
\centering
  \begin{subfigure}[b]{0.8\textwidth}
    \centering
    \includegraphics[width=\textwidth]{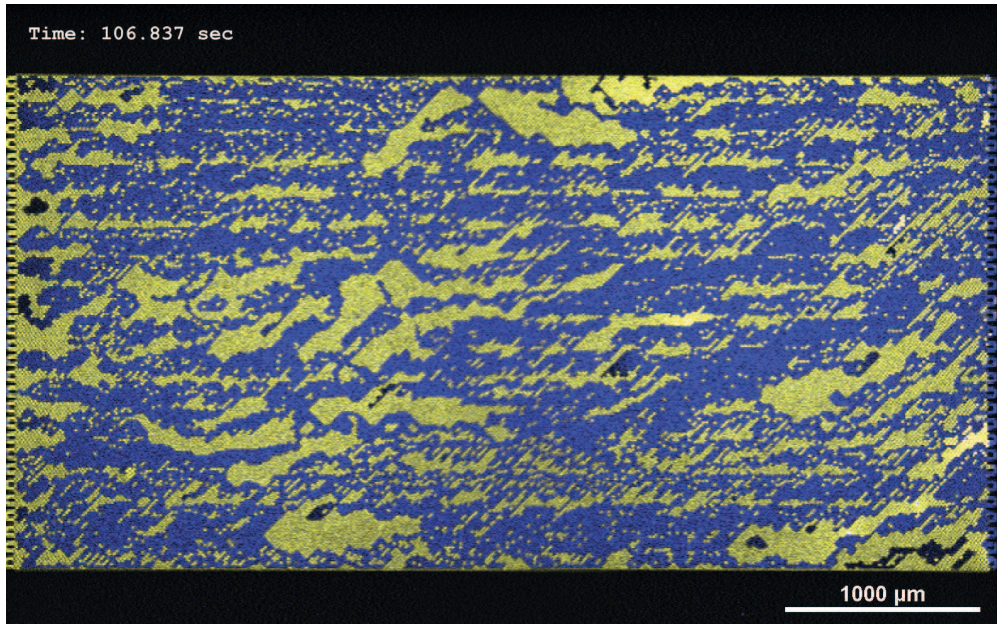}
    \caption{}
    \label{fig:ConfocalResults_LastFrame}
  \end{subfigure}
  \hskip 2ex
  \begin{subfigure}[b]{0.8\textwidth}
    \centering
    \includegraphics[width=\textwidth]{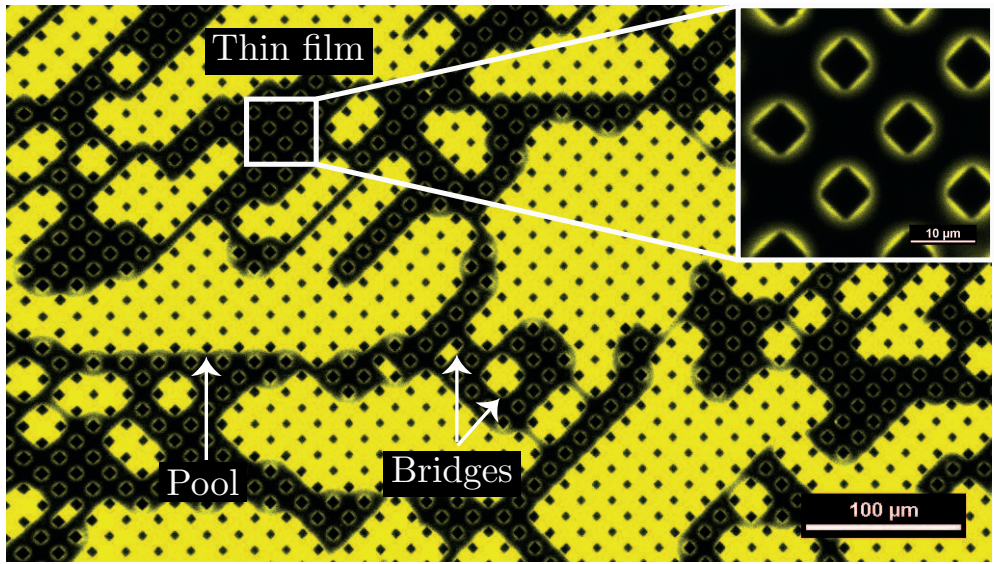}
    \caption{}
    \label{fig:Structures_of_wetting_film_new}
  \end{subfigure}
  \caption{Experiment using laser scanning confocal microscope (LSCM). (a) Image after water transport through the liquid infused chip at $Q$=0.2 $\mu$l/s ($Ca$=1.23$\times$10$^{-4}$) (yellow color is the oil phase and blue color is the water phase). (b) Image at 20x magnification showing different configurations of the residual wetting fluid.}
  \label{fig:Fig7}
\end{figure}

A typical result of the microfluidic displacement is shown in Figure \ref{fig:Fig7}. Figure \ref{fig:ConfocalResults_LastFrame} displays the last time frame of the corresponding movie (see SI for the movie). In this figure, yellow and blue colors correspond to the infused liquid (SO AR20) and water, respectively. A highly ramified pattern with trapped infusion liquid is observed (Figure \ref{fig:Structures_of_wetting_film_new}). Based on the used capillary number ($Ca\simeq10^{-4}$ see Table \ref{tab:3}) and viscosity ratio ($M=0.05$), the displacement pattern corresponds to the capillary fingering invasion regime in drainage. Fingers in capillary fingering show spreading across considerable part of the chip and growth is seen in all directions. When capillary fingering occurs some wetting fluid becomes entrapped due to the complex displacement pattern \cite{doi:10.1021/ef101732k}. The residual wetting fluid can be observed as pools, bridges and thin films around the pillars (Figure \ref{fig:Structures_of_wetting_film_new}). Observation of thin films around pillars (insert of Figure \ref{fig:Structures_of_wetting_film_new}) further confirms liquid-lining after displacing with water (Figures \ref{fig:SLIM_pore_white_new} and S6). The presence of this liquid layer is crucial for any potential anti-fouling properties of liquid-infused surfaces due to minimum contact of foulants with solid material of the surface \cite{837566}.

\begin{table}
\small
\centering
\caption{Irreducible water saturation values (S$_{\text{nwr}}$) for the microfluidic experiment done at different flow rates.} \label{tab:3}
\begin{tabular}{M{3.5cm} M{5.5cm} M{2.5cm} M{2cm}}
\hline
Microfluidic Q ($\mu$l/s) & Membrane flux (m$^3$/m$^2$ s) & $Ca$ & S$_{\text{nwr}}$ \\
\hline
0.2 & 0.0013 & 1.23$\times$10$^{-4}$ & 0.56\\
0.4 & 0.0026 & 2.46$\times$10$^{-4}$ & 0.62\\
0.8 & 0.0053 & 4.91$\times$10$^{-4}$ & 0.63\\
1.6 & 0.011 & 9.28$\times$10$^{-4}$ & 0.69\\
2.9 & 0.018 & 1.72$\times$10$^{-3}$ & 0.73\\
\hline
\end{tabular}
\end{table}

\begin{figure}
\centering
\includegraphics[width=0.7\linewidth]{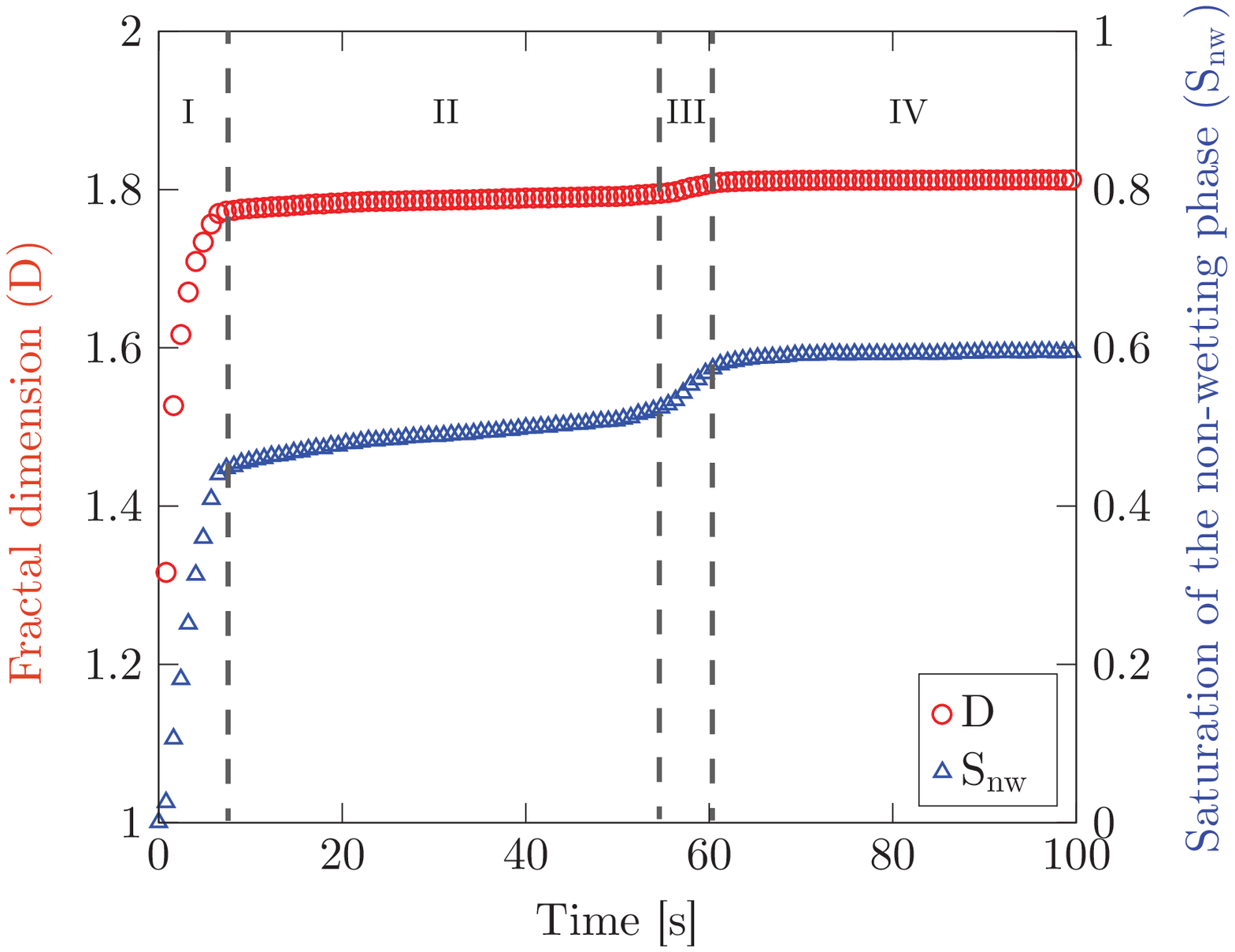}
\caption{Total Saturation of the non-wetting phase (S$_{\text{nw}}$) and fractal dimension ($D$) as a function of time for the corresponding displacement movie. During phase (I) formation of the fingers takes place. During phase (II) the saturation profile is evolving. Phase (III) corresponds to the breakthrough point and phase (IV) corresponds to the steady-state regime where the saturation profile is preserved.}
\label{fig:S_nw_and_Fractal_analysis_2_thicker_without_colors}
\end{figure}

For quantitative description of the observed patterns and better understanding of the displacement mechanism, the patterns are analyzed in terms of total saturation S$_{\text{nw}}$ of the non-wetting phase (water in this case), the local saturation (and their evolution over time) and the fractal dimension $D$. The total saturation S$_{\text{nw}}$ is the ratio of the total area of the non-wetting phase to the total available area of the chip \cite {arXiv:0909.0758, doi:10.1021/ef101732k}. For measuring the local saturation, each image is divided vertically into 33 slices and S$_{\text{nw}}$ is calculated for each slice. The MATLAB bio-format toolbox is used for image processing \cite{10.7717/peerj.2674}.

The fractal dimension $D$ is given by

\begin{equation}
D = \frac{\log B}{\log a}. \label{eq7}
\end{equation}

\noindent Here, $B$ is the number of boxes that cover the pattern and $a$ is the magnification factor which is the inverse of the box size.
A box counting algorithm is used to calculate $D$ ({\href{https://nl.mathworks.com/matlabcentral/fileexchange/30329-hausdorff--box-counting--fractal-dimension}{Hausdorff (Box-Counting) fractal dimension code}}) (See SI for more details). $D$ for this kind of analysis is a non-integer value which ranges between 1 and 2. $D$=1 for a straight line and $D$=2 for a fractal pattern which completely fills up a 2D plane \cite{JensFederFractal}.

Figure \ref{fig:S_nw_and_Fractal_analysis_2_thicker_without_colors} shows the total saturation S$_{\text{nw}}$ and the fractal dimension $D$ as a function of time. There are four distinct phases in the total saturation plot. In phase I, a sharp increase is observed corresponding to the formation of the fingers. Phase II corresponds to the evolution of the formed fingers before the breakthrough point. During phase III the breakthrough point has been reached. At this point a decrease in specific interfacial length for the boundary between two phases is observed corresponds to the removal of the larger mobile oil blobs (Figure S5(b)). In phase IV, S$_{\text{nw}}$ reaches a constant water saturation value of 0.6 after the breakthrough point where the oil blobs are completely immobile. The corresponding saturation value for the displacing liquid is the saturation at which the displaced liquid goes from being mobile to being immobile, i.e. irreducible saturation (S$_{\text{nwr}}$).

The irreducible saturation values (S$_{\text{nwr}}$) of other experiments done at higher flow rates are shown in Table \ref{tab:3}. The results show that at the highest flow rate which corresponds to the highest flux value of the membrane experiment ($1.8\times10^{-2}$ m$^3$/m$^2$ s), 27\% of the infused liquid will remain in the chip. This is lower than the amount of remaining infused liquid in the membrane experiments (Table \ref{tab:2}). The reason can be attributed to the presence of interconnected pores and irregular shaped pores in the membrane where more infused liquid can be trapped.

The fractal dimension $D$ reaches its steady state value ($D_s$) of 1.8 at the breakthrough point. This confirms that the displacement mechanism falls within the flow regime of invasion percolation with trapping (IPT), a statistical model which is used to describe capillary fingering. It is established that IPT patterns can be identified with fractal dimension between 1.8 and 1.83 \cite{lenormand_touboul_zarcone_1988, Lenormand1989}.

\begin{figure}
\centering
\includegraphics[width=0.7\linewidth]{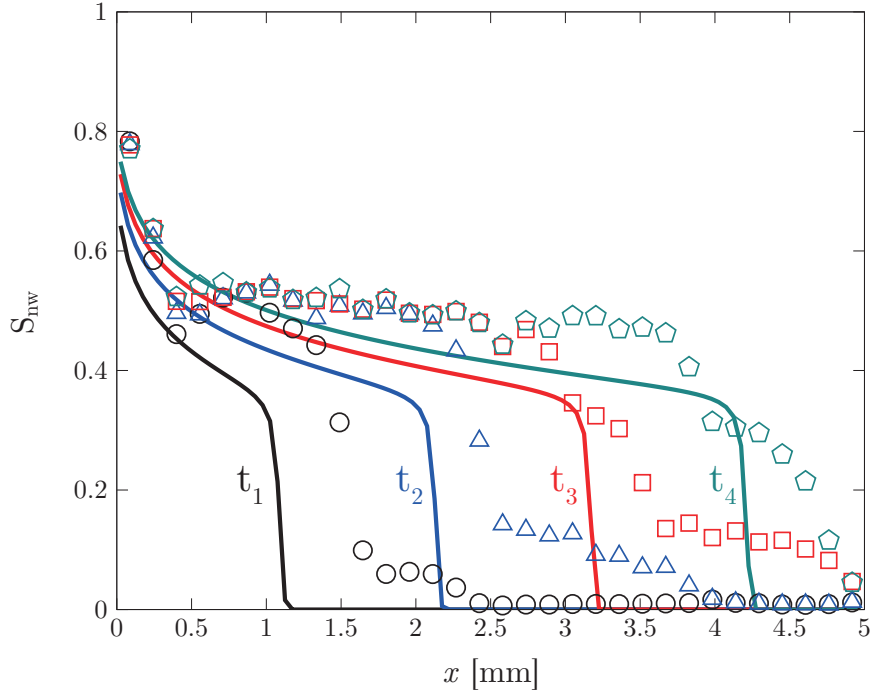}
\caption{Buckley-Leverett experimental results (symbols) and modelling results (solid line) of the shock front movement for four different times, i.e., t$_\text{1}$ = 1 s (black), t$_\text{2}$ = 2 s (blue), t$_\text{3}$ = 3 s (red), and t$_\text{4}$ = 4 s (green).}
\label{fig:Buckley_Leverett_thicker_new_empty}
\end{figure}

In order to better quantify the formation of the fingers in phase I, local saturation of the non-wetting phase is plotted as a function of time and location (see Figure S8). Local saturation provides a detailed description of the dynamics of liquid-liquid displacement during the invasion. The initial formation of the fingers corresponds to the movement of a shock front toward the outlet (Figure \ref{fig:Buckley_Leverett_thicker_new_empty}). In order to model the displacement behaviour and movement of the shock front, the Buckley-Leverett model is used \cite{Buckley1942Leverett},

\begin{equation}
\frac{\phi}{u}\frac{\partial S_{\text{nw}}}{\partial t}+\frac{df_{\text{nw}}}{dS_{\text{nw}}}\frac{\partial S_{\text{nw}}}{\partial x}=0. \label{eq8}
\end{equation}

\noindent Here $\phi$ is the porosity, and $u$ is the total fluid velocity in the direction of the flow. $f_{\text{nw}}$ is fractional flow of the non-wetting phase, which is defined as the ratio of the non-wetting phase velocity to the total velocity, i.e., $f_{\text{nw}}=u_{\text{nw}}/u$. This model is the best known analytical approach for investigation and modelling of two-phase flow in porous media \cite{Pinder2008Essentials,Sahimi2011Flow,marle1981multiphase} (see SI for derivation and solution details).

The modelling result is shown in Figure \ref{fig:Buckley_Leverett_thicker_new_empty}. The model can predict the experimental results for the movement of the shock front. The differences may originate from the simplifications in the model as well as experimental accuracy. The displacement observations using LSCM is done at 4x magnification and each time frame of the corresponding movie is divided to 33 slices to obtain the local saturation plot. This is the smallest achievable slice size in order to avoid noticeable experimental noise.

\section{Conclusions}
\label{s:conc}
We have reported on the retention of the infusion liquid in slippery liquid infused membranes (SLIM) during water permeation and microscopic observation of the displacement mechanism. The membrane experiments have been done via liquid-liquid displacement porometry (LLDP) by pushing pure water through SLIM in a flux-controlled mode. The pressure-flux results confirmed the presence of remaining infusion liquid (43\%) after displacing with water. Comparison of different liquid-infused membranes with a dry one further indicates the presence of liquid-lined pores. Infused pores were displaced according to the corresponding capillary pressure and remained open once flow was stopped. This suggested the preferential flow path ways for water transport through the membranes which corresponds to the capillary fingering displacement regime. The related physics regarding two-phase flow in porous media was used to confirm the observed displacement mechanism in SLIMs. In order to better understand the physical displacement process, a microfluidic chip was used for direct visualization using laser scanning confocal microscope (LSCM). A highly ramified structure with trapped infusion liquid was observed. Residual wetting structures were observed in the form of pools, bridges, and thin films around pillars. The presence of thin films further confirmed liquid-lining after displacing with water. Fractal breakthrough patterns were analyzed in terms of total and local saturation (S$_\text{nw}$) and fractal dimension ($D$). The saturation profiles indicated the dynamic distribution of the infusion liquid during the displacement process. Local saturation provides details on the formation of the fingers which corresponds to the movement of a shock front toward the outlet. Buckley-Leverett model was used to predict the experimental results of the shock front movement. The observed patterns along with the fractal analysis confirmed that the experiment falls within the flow regime of capillary fingering which can be described by an invasion percolation with trapping (IPT) model. The saturation of the water phase reached the steady-state value of 0.73 at the same capillary number as membrane experiments corresponds to the highest flux value. This showed that 27\% of the oil still remains in the chip. This study showed the retention of liquid-lining under cyclic pressure-flux testing during immiscible displacement process. The presence of the liquid-lined pores after displacement with water is crucial for anti-fouling characteristics of SLIM, which makes it a potential candidate for separation processes.

\section{Materials and Methods} \label{s:matmethods}
\subsection{Materials}
PVDF (Solef 6020/1001) was received from Solvay Solexis, France. 1-methyl-2-pyrrolidinone (NMP) (99\% extra pure) was purchased from Acros Organics, The Netherlands. Ethanol (99.8\%) was supplied from Atlas and Assink chemical company, The Netherlands. Krytox GPL oil 101 was purchased from MAVOM chemical industry, The Netherlands. Silicone oil AR20 was purchased from Sigma-Aldrich, The Netherlands. Trichloro(1H,1H,2H,2H-perfluorooctyl)silane (FOTS, 97\%) was purchased from Sigma-Aldrich, The Netherlands. Perylene (sublimed grade, 99.5\%) and Rhodamine 6G (dye content 99\%) as fluorescent dyes for oil and water phases respectively were purchased from Sigma-Aldrich, The Netherlands. n-Hexane and 2-propanol (analysis grade) were purchased from Merck milipore, The Netherlands. Sulfuric acid (H$_2$SO$_4$ 95-98\%) was purchased from Merck milipore, The Netherlands. Hydrogen peroxide (H$_2$O$_2$) solution 30\% (w/w) in H$_2$O was supplied from Sigma-Aldrich, The Netherlands.

\subsection{Membrane fabrication}
\label{subs:memfab}
The polymer solution was prepared by mixing 15 wt.\% PVDF in NMP with a mechanical stirrer overnight at 80$^{\circ}$C. The solution was cast on a glass plate using a casting knife at an initial thickness of 500 $\mu m$. The cast membrane was immediately submerged in water/NMP (30:70 vol.\%) as the coagulation bath for 60 min. To remove the remaining NMP from the membranes, they were kept in ethanol subsequently for another 60 min. The films were then taped to a piece of paper to prevent curling and left to dry in fume hood (60 min) before placing them in a 30$^{\circ}$C vacuum oven overnight.

\subsection{Fabrication of dense PVDF}
\label{subs:Fabdense}
Dense PVDF was made by casting the polymer dope solution on a glass plate using the same procedure as described for membrane fabrication. The cast polymer solution was placed in a box and dried with a flow of nitrogen for two days.

\subsection{Fabrication of SLIM and pre-wet membrane} \label{subs:FabSLIMPreWet}
An overcoat layer (15.5 $\mu$l cm$^{-2}$) of the low surface tension liquid (Krytox 101) was added to the membranes using a micropipette. The liquid spontaneously infiltrated the pores via capillary wicking. The samples were further placed vertical (2-3 h) for gravity-induced removal of the excess liquid.

The pre-wet membrane was prepared using the same procedure by adding ethanol to the membrane. To replace ethanol with water in membrane pores, the sample was placed in a beaker of water (200 $m$l) and left overnight.

\subsection{Membrane characterization}
\label{subs:Memchara}
The membrane was characterized using contact angle Goniometer (Dataphysics OCA20), capillary flow porometer (Porolux-1000), and scanning electron microscope (SEM) (JEOL 5600 LV). See SI for details on the techniques.

\subsection{Laser scanning confocal microscopy (LSCM) experiments}
\label{subs:LSCM}
The liquid-liquid displacement in the microfluidic chip is observed using an inverted laser scanning confocal microscope (LSCM) (A1 system, Nikon Corporation, Japan) with a 4x dry objective (CFI Plan Fluor 4x/0.13, numerical aperture (NA) = 0.13, working distance (WD) = 17.2 mm). The scanning area is chosen the same as the area of the pillar structure in the microfluidic chip, i.e. $5\times2.5$ mm$^2$.

\subsection{Hydrophobization of the microfluidic chip}
\label{subs:hydrophob}
In order to render the microfluidic chip hydrophobic, it was hydrophobized using deposition of trichloro(1H,1H,2H,2H-perfluorooctyl)silane (FOTS) via vapor-induced method. To achieve this, a glass bottle with gas inlet and outlet was filled with 12 $\mu$L of FOTS. The inlet was connected to nitrogen gas and the outlet was connected to the chip using a chip holder. The whole set-up was then placed in oven at temperature of 100$^{\circ}$C. The outflow from the chip was directed to a beaker containing water to neutralize and absorb the permeate gas. The acidity of water was checked to further ensure the presence of FOTS vapor in the chip via formation of hydrochloric acid (HCl) in water. Before hydrophobization, the chip was cleaned with piranha solution (H$_2$SO$_4$:H$_2$O$_2$ (3:1) vol.) and rinsed with pure water. Since the chip should be completely dry before hydrophobization, water was replaced with 2-propanol and then n-hexane. The chip was dried in oven (temperature of 100$^{\circ}$C) overnight while having a flow of nitrogen gas through the channels.

\section{ACKNOWLEDGMENTS}
We thank Gabriela Berenice Diaz Cortes from Delft University of Technology, The Netherlands for invaluable advice on Buckley-Leverett analysis using MATLAB reservoir simulation tool box and Yanshen Li, guest PhD at Physics of Fluids (POF) group at University of Twente for help on laser scanning confocal microscope. We also thank Jan van Nieuwkasteele and BIOS lab-on-a-chip group at Univerity of Twente for the design and fabrication of the microfluidic chip. This work was performed in the cooperation framework of Wetsus, European Centre of Excellence for Sustainable Water Technology (www.wetsus.nl). Wetsus is co-funded by the Dutch Ministry of Economic Affairs and Ministry of Infrastructure and Environment, the European Union Regional Development Fund, the Province of Frysl\^{a}n, the Northern Netherlands Provinces. This work is part of a project that has received funding from the European Union's Horizon 2020 research and innovation programme under the Marie Sk\l{}odowska-Curie grant agreement No. 665874. This work was also supported by the Netherlands Center for Multiscale Catalytic Energy Conversion (MCEC), an NWO Gravitation program funded by the Ministry of Education, Culture and Science of the government of the Netherlands. This work is part of the Vici project STW 016.160.312 which is financed by the Netherlands Organisation for Scientific Research (NWO).

\bibliography{rsc}
\bibliographystyle{rsc}

\end{document}